\documentclass[twocolumn,prl,aps,flushbottom,showpacs,10pt]{revtex4-1}
\usepackage{xspace,amsmath,amsfonts,amsthm,amssymb,amsbsy,graphicx,color}

\begin{document}

\newtheorem{theo}{Theorem} \newtheorem{lemma}{Lemma}

\title{Simulating many-body lattice systems on a single nano-mechanical resonator} 

\author{Kurt Jacobs}

\affiliation{ 
Advanced Science Institute, RIKEN, Wako-shi 351-0198, Japan \\ 
Department of Physics, University of Massachusetts at Boston, Boston, MA 02125, USA \\
Hearne Institute for Theoretical Physics, Louisiana State University, Baton Rouge, LA 70803, USA
} 

\begin{abstract} 
We show that lattice systems, such as the Bose-Hubbard model, can be simulated on a single nano- or micro-mechanical resonator, by exploiting its many modes. The on-site Hamiltonians are engineered by coupling the mechanical modes to the modes of a pair of optical or stripline resonators, and the connections between the lattice sites are engineered in a similar way. The lattice network structure is encoded in the frequency components of the fields driving the resonators. This three-resonator configuration also allows universal quantum computing on the nano-resonator. 
\end{abstract}



\maketitle 

Current methods to simulate many-body lattice systems using engineered mesoscopic systems achieve this by fabricating a lattice of the latter~\cite{Hartmann06, Angelakis07, Carusotto09, Nunnenkamp11, Houck12}. Here we show how to create an effective many-body lattice using a total of three mesoscopic resonators. The bodies, or \textit{lattice sites}, are the multiple modes of a nano-mechanical resonator~\cite{Blencowe04, Aspelmeyer12}. The nonlinear Hamiltonian for each mode is engineered by coupling the nano-resonator to a pair of optical or superconducting resonators: pairs of the auxiliary modes create the nonlinearity for each mechanical mode. The interactions (links) between lattice sites are also obtained by a coupling to the modes of an auxiliary resonator, which may be one of the previous pair, or a third auxiliary. Each lattice link is created by one mode of this auxiliary, the structure of the lattice being determined by the field that drive it: each frequency component in the driving field selects which nano-mechanical mode (or ``nano-mode'') is coupled to which auxiliary mode, and thus which nano-modes are coupled together. This configuration dramatically reduces the number of mesoscopic systems required to simulate a many-body system, and also means that the lattice network can be ``reprogrammed'' by changing the driving field. 

Here we focus on constructing the Bose-Hubbard (BH) model~\cite{Fisher89}, but we also discuss other lattice Hamiltonians that could be engineered in a similar way, including spin lattices. The BH model is especially interesting because of it's simplicity, and the fact that it possesses both Mott-insulator and superfluid phases. It consists of a chain of oscillators, with nearest-neighbor linear coupling, each with a Kerr nonlinearity. The Hamiltonian for the BH model is 
\begin{equation}
   H_{\mbox{\scriptsize BH}} = \sum_n \xi (a_n^\dagger a_n)^2 + \sum_n \zeta (a_n a_{n+1}^\dagger + a_n^\dagger a_{n+1}) , 
\end{equation}
where $\{a_n\}$ are the mode annihilation operators.  

In recent years it has been shown that the modes of a doubly-clamped nano-resonator are intrinsically coupled via a nonlinear interaction, although very weakly~\cite{Westra10}. While this coupling is intriguing, as it can be enhanced by driving and exploited in various ways~\cite{Mahboob12, Mahboob11}, it is still too weak in present systems for our purposes, since we wish to work down to the single-phonon level. To engineer a desired lattice Hamiltonian connecting the modes of a nano-resonator, we must therefore couple the resonator to another system. Nonlinearities for a single nano-mode, and coupling between nano-modes can be engineered by coupling to a non-linear system (see, e.g.~\cite{Beltran07, Boissonneault08, Mallet09, Jacobs09c, Leib12}) such as a transmon qubit~\cite{You07, Koch07}. The problem with this technique is that if we couple all the modes to a single qubit we cannot avoid coupling all the modes together simultaneously. While creating an all-to-all network is an interesting prospect in itself, the above procedure can only couple the nano-modes in a lattice configuration by using many individually fabricated qubits. We instead create a lattice by using the multiple (linear) modes of a single auxiliary resonator as the set of auxiliary systems to create the lattice links. Then a method discovered recently by Ludwig \textit{et al.}~\cite{Ludwig12} and Stannigel \textit{et al.}~\cite{Stannigel12}, building on recent experimental and theoretical work~\cite{Grudinin10}, allows us to use these linear modes to create the required nonlinearity for each of the nano-modes (with the addition of a second auxiliary resonator). The method works by coupling a single nano-mode to two auxiliary modes with the same frequency, while linearly mixing the auxiliary modes. There are many ways to linearly mix modes, and examples include optical modes coupled by a polarization rotator; optical toroidal resonators coupled via an evanescent field~\cite{Grudinin10}; or  superconducting resonators coupled electrically~\cite{Bajjani11}. As we elucidate, each pair of mixed modes acts effectively like a single qubit, and it is the coupling to these qubits that allows us to create the nonlinearities~\cite{Jacobs09c}. 

For notational simplicity, all our Hamiltonians will be scaled by $1/\hbar$, so that they have units of inverse seconds (or equivalently rad/s). With this scaling, all the rate-constants that appear in the Hamiltonian are also the rate-constants that appear in the resulting equations of motion. We therefore write the Hamiltonian for a nano-mechanical resonator as $H_{\mbox{\scriptsize m}} = \sum_{n} \omega_n a_n^\dagger a_n$, where $\omega_n$ and $a_n$ are the respective frequency and annihilation operator for each mode. The frequency spectrum is usually highly complex, and some modes will have much higher quality factors than others. This is not crucial for our purposes, however, as we will be able to select the modes we use by choosing the driving frequencies. 

We must also take into account that all our resonators are subject to damping. We describe this damping using the standard Markovian master equation in the Lindblad form~\cite{Breuer07}. Each source of damping or decoherence is described by including a term $k \mathcal{D}(A) \rho$ in the equation of motion for the density matrix $\rho$, where $\mathcal{D}(A) \rho = A \rho A^\dagger - (1/2)(A^\dagger A \rho + \rho A^\dagger A )$~\cite{Wiseman93}. The rate constant $k$ is called the damping rate, and the dimensionless operator $A$ is called the Lindblad operator. The Lindblad operator for the damping of each mode is simply the annihilation operator for that mode, and their respective damping rates will be defined as needed. 

We first consider how to couple selected mechanical modes together using the modes of an auxiliary (optical or microwave) resonator. If we denote the frequencies and annihilation operators of the auxiliary modes by $\Omega_j$ and $b_j$, then the auxiliary Hamiltonian is $H_{\mbox{\scriptsize aux}} = \sum_j \Omega_j b_j^\dagger b_j$, and the basic coupling to the nano-resonator is~\cite{Law95} 
\begin{equation}
   H_1 = \sum_{nj} g_{nj} b_j^\dagger b_j (a_n + a_n^\dagger) , 
   \label{basicint}
\end{equation}
in which every mechanical mode is coupled to every auxiliary mode by some rate-constant $g_{nj}$. We now create an effective linear interaction by using the standard method employed in sideband cooling --- driving the auxiliary with a coherent field --- but we modulate this drive so that it contains $N$ frequencies $\nu_k$, $k =1,\ldots,N$. The auxiliary field will settle down to some steady-state, given by a coherent part plus quantum fluctuations. Denoting the coherent part by $\beta(t) = \sum_k \beta_k \exp(-i\nu_k t)$, we remove it by transforming to the ``displacement picture''~\cite{Wiseman93}. The interaction Hamiltonian becomes 
\begin{equation}
   H_1' = \sum_{nj} g_{nj} [ b_j^\dagger + \beta^*(t) ] [ b_j + \beta(t) ] (a_n + a_n^\dagger) . 
   \label{H1p}
\end{equation}
We now move to the interaction picture with respect to the free Hamiltonians of the two resonators, and perform the rotating-wave approximation (RWA)~\cite{Irish07}. It is this that selects which modes will interact, by eliminating pairs that do not satisfy the three-way resonance condition $\omega_n + \Omega_j - v_{k} = 0$ between the modes and the driving frequencies. The result is $\tilde{H}_{\mbox{\scriptsize int1}} = \sum_k \tilde{H}_k$, with 
\begin{equation} 
   \tilde{H}_k = \sum_{k} g_{k} |\beta_k| ( a_k b_k^\dagger + b_k a_k^\dagger ) , 
\end{equation} 
where each $k$ labels a pair of modes $(a_k,b_k)$. Note that the two additional terms multiplying $(a_n + a_n^\dagger)$ in Eq.(\ref{H1p}), namely $b_k^\dagger b_k + |\beta_k|^2$, have averaged out under the RWA applied with respect to the mechanical frequency.

We now select the driving frequencies so that two mechanical modes, $a_k^{(1)}, a_k^{(2)}$ are coupled to the same auxiliary mode $b_k$, and then adiabatically eliminate the auxiliary. To perform this adiabatic elimination we first detune the nano-modes from resonance with the auxiliary by an amount $\Delta \omega$. The adiabatic elimination requires that this detuning and the damping rate of the auxiliary, $\kappa$, are much faster than the dynamics of the mechanical modes, which in our interaction picture are given purely by the coupling rate $g_k$. We can now set $\dot{b}_k = 0$, and this ``slaves'' the state of the auxiliary to that of the nano-modes so that $b_k = g_k \beta_k (a_k^{(1)} + a_k^{(2)})/(\kappa - i\Delta \omega) + \tilde{n}(t)$. Here $\tilde{n}(t)$ is quantum input noise that comes from the damping~\cite{Dunningham97}. After applying the RWA once more, we obtain the simple linear coupling  
\begin{equation} 
   \tilde{H}_k = \sum_{k} \left[ \frac{i (g_k \beta_k)^2 \Delta \omega}{\Delta \omega^2 + \kappa^2} \right] ( a_k^{(1)} a_k^{(2)\dagger} - a_k^{(2)} a_k^{(1)\dagger} ) . 
   \label{Heff1}
\end{equation} 
Since the auxiliary mode is damped, when eliminated it also induces an effective damping in the mechanical modes. The effective damping rate is $\gamma_k' = (g_k \beta_k)^2 \kappa (\Delta \omega^2 + \kappa^2)$, and the Lindblad operator is $A_k = a_k^{(1)} + a_k^{(2)}$. Note that since we can couple any number of nano-modes to a single auxiliary mode, we can use the above technique to couple any subset of lattice sites in an all-to-all network if we wish. 

Implementing the above coupling procedure gives us a many-body lattice of harmonic oscillators (or bosons). To make the lattice really interesting we need to add a nonlinearity to each oscillator. As mentioned above, we do this using the technique in~\cite{Ludwig12, Stannigel12}. We couple each mechanical mode to a pair of linearly coupled auxiliary modes of the same frequency, whose annihilation operators we denote by $\tilde{c}_k$ and $\tilde{d}_k$. The coupling Hamiltonian is the same as that in Eq.(\ref{basicint}), except that the auxiliary modes now come in pairs, and the Hamiltonian contains linear mixing of these pairs. The interactions between all the modes are therefore given by  
\begin{equation}
   H_2 = \sum_{nj} f_{nj} (\tilde{c}_j^\dagger \tilde{c}_j + \tilde{d}_j^\dagger \tilde{d}_j ) (a_n + a_n^\dagger )  + s_j (\tilde{c}_j \tilde{d}_j^\dagger + \tilde{d}_j \tilde{c}_j^\dagger) ,    
\end{equation}
where $s_j$ is the rate of mixing for each linear pair. To begin we rewrite the full Hamiltonian in terms of the transformed modes $c_j \equiv (\tilde{c}_j + \tilde{d}_j)/\sqrt{2}$ and $d_j = (\tilde{c}_j -\tilde{d}_j)/\sqrt{2}$ (these are referred to as ``supermodes''~\cite{Grudinin10}), and the result is  
\begin{eqnarray}
   H & = & \sum_{n} \omega_n a_n^\dagger a_n + \sum_j  \left[ (\Omega_j + s_j) c_j^\dagger c_j  + (\Omega_j - s_j) d_j^\dagger d_j  \right]  \nonumber \\ 
   & & + \sum_n f_{nj} (d_j  c_j^\dagger + c_j d_j^\dagger ) (a_n + a_n^\dagger )  . 
   \label{3way}
\end{eqnarray}
This transformation to the new modes $c$ and $d$ is revealing. It shows us that we have two auxiliary modes now at two different frequencies ($\Omega_j \pm s_j$), and that the non-linear coupling to the nano-modes flips photon between these two modes at a rate proportional to $\langle a_n + a_n^\dagger\rangle$~\footnote{This three-mode coupling can also be obtained in a membrane-in-the-middle cavity~\cite{Miao09}}. Further, the interaction has a clear resonance condition: in the interaction picture the term $ f_{nj} d_j  c_j^\dagger a_n$ and its Hermitian conjugate will be constant if $2s_j = \omega_n$. By varying the mode mixing rate $s_j$ across the modes, a single auxiliary mode can be selected to interact with a single mechanical mode. If one is able to modulate the non-linear mechanical/electrical coupling rate $f_{nj}$ at a frequency $\delta_{nj}$, then the resonance condition becomes $2s_j = \omega_n + \delta_{nj}$, providing a second means of tuning the mode selection.    

Now focus on the interaction between a single nano-mode and auxiliary mode-pair, where we have detuned the nano-mode from the resonance by $\Delta$ so that $2s_j = \omega_n + \Delta$. Assuming that all other nano-modes are much further from resonance with the mode-pair than the one we have selected, applying the RWA to the Hamiltonian above, and moving into the interaction picture with respect to the nano-mode frequency $\omega_n$, we have  
\begin{eqnarray}
   H' & = & \Omega (c^\dagger \! c  + d^\dagger \! d ) + \Delta (c^\dagger \! c  - d^\dagger \! d ) + f [a dc^\dagger + (a d) \! ^\dagger \! c ] \nonumber .  
\end{eqnarray}
We note that since $c^\dagger \! c  + d^\dagger \! d$ commutes with all the other operators in $H'$, the high frequency $\Omega$ has no effect on the interaction dynamics, and thus on the perturbation analysis that we will shortly use. But it will be important when we adiabatically eliminate the cavity mode $c$.  The  Hamiltonian $H'$ is especially interesting because the operators  $dc^\dagger$ and $c^\dagger \! c  - d^\dagger \! d$ have the commutation relations of spin operators --- they are the Schwinger representation of angular momentum, with identification $\sigma_z = c^\dagger \! c  - d^\dagger \! d$ and $\sigma_+ = dc^\dagger$. The linear coupling of the modes has converted the resonator's basic nonlinear interaction with the linear modes into a linear interaction with a nonlinear system. Because the nano-resonator mode is now effectively coupled to a spin-1/2 system, via a spin operator that does not commute with the spin Hamiltonian (in this case $\sigma_z$), it is possible at least in theory to engineer a range of non-linear Hamiltonians for the mode~\cite{Jacobs09c}. 

The Kerr (or $\chi^{(3)}$) nonlinearity, can now be created in at least two ways. We can make $\Delta$ sufficiently large ($\Delta \gg \omega_m, f$) that the interaction is a perturbation for the qubit. If we use time-independent perturbation theory to diagonalize the Hamiltonian in powers of $\varepsilon \equiv (f/\Delta)$, the even powers give the effective interaction $\sigma_m(a^\dagger \! a)^{2m}$, $m=1,2,\ldots$, where $\sigma_m$ is some Pauli spin-operator for the qubit. The forth-order term ($m=2$) is the Kerr term. However, to create the Kerr in this way requires that we prepare the qubit in the correct eigenstate of $\sigma_m$, and this is not so easy for our effective optical qubit. The other method, used in~\cite{Ludwig12, Stannigel12}, is to expand the perturbative series to second order, giving the interaction $H_{\mbox{\scriptsize eff}} = \sigma_1(a^\dagger \! a)$, and then adiabatically eliminate the qubit. This is essentially the same adiabatic elimination procedure we used above. In this case the qubit operator $\sigma_1$ becomes effectively $(a^\dagger \! a)$, so that $H_{\mbox{\scriptsize eff}}$ becomes the Kerr Hamiltonian. 

To perform the perturbative diagonalization to second-order, we transform the Hamiltonian using $U = \exp[-i(a dc^\dagger - a^\dagger \! d^\dagger \! c)/2]$. The resulting effective Hamiltonian commutes with both $c^\dagger \! c$ and $d^\dagger \! d$, so it has the same eigenstates as the latter. Leaving mode $d$ in the vacuum state, all appearances of $d^\dagger \! d$ disappear, and we are left with the very simple non-linear Hamiltonian   
\begin{eqnarray}
   H'' & = & \Omega \, c^\dagger \! c + \varepsilon \left( \frac{ f}{4} \right) c^\dagger \! c \, a^\dagger \! a  . 
\end{eqnarray}
Due to the damping of the auxiliary modes at the rate $\kappa$, the transformation $U$ also creates an effective damping for the resonator mode at rate $\gamma' = \varepsilon^2 \kappa /4$, with Lindblad operator $L_{\gamma'} = c a$.  

We now drive mode $c$ with a field $\alpha$, shift to the displacement picture to create the effective interaction $\alpha (c + c^\dagger) (a^\dagger \! a)$ (the same procedure that gives Eq.(\ref{H1p}) above), and adiabatically eliminate the auxiliary mode $c$. It is now useful to remember the following general rule: given a mode $b$ that interacts with another system via $(b + b^\dagger) K$, where $b$ has a fast oscillation rate $\Omega$ and a fast damping rate $\kappa$, then to second-order in the adiabatic elimination, $\Omega$ produces the effective Hamiltonian $K^2$ (in a manner similar to perturbation theory), and the damping $\kappa$ induces an effective damping (or continuous measurement) via the Lindblad operator $K$.  For the above system the result is the Hamiltonian~\cite{Stannigel12}
\begin{eqnarray}
   H_{\mbox{\scriptsize Kerr}} & = & \left[ \frac{(\varepsilon f \alpha /4)^2 \Omega}{\Omega^2 + \kappa^2} \right] (a^\dagger \! a)^2 , 
      \label{Heff2}
\end{eqnarray}
with decoherence at rate $\Gamma = (\varepsilon f \alpha/4)^2 \kappa/(\Omega^2 + \kappa^2)$ via the Lindblad operator $L = a^\dagger \! a$.

The two effective Hamiltonians we have described, Eqs.(\ref{Heff1}) and (\ref{Heff2}) allow the Bose-Hubbard chain to be created in a single resonator. It is interesting to consider what other many-body systems might be engineered in a similar way, and whether one could use an optical rather than a mechanical resonator to hold the resulting lattice. To address the latter question first, we note that the asymmetry between the mechanical and optical/superconducting resonators is that given by the interaction in Eq.(\ref{basicint}). All we have to do to generate a Kerr non-linearity for the cavity is to adiabatically eliminate the nano-resonator. This is the basis of the schemes in~\cite{Bose97, Mancini97}. However, to form a lattice we need to selectively couple only one nano-mode to each optical mode, and do this for a number of optical modes. This would be relatively simple if one could modulate the coupling strength $g_n$, which might be possible, for example, using a membrane-in-the-middle scenario~\cite{Thompson08}. Once we have created a Kerr Hamiltonian for each mode, we can selectively couple the modes together using the technique described above for nano-resonators. 

The main limitation in constructing lattice Hamiltonians stems from the need to create higher-order nonlinearities perturbatively. Because of this, the higher the non-linearity the weaker its strength. Nevertheless, the third-order ($\chi^{(3)}$) nonlinearities we have considered here do give further interesting possibilities. As shown in~\cite{Stannigel12}, the cross-Kerr coupling, $a^\dagger \! a \, b^\dagger \! b$, between two modes $a$ and $b$ can be engineered in precisely the same way as the Kerr for a single mode, by coupling two nano-modes to the same pair of optical modes. Another intriguing possibility is to use the three-way coupling between two super-modes and the nano-mode in Eq.~\ref{3way}, to simulate a lattice with three-body interactions. Finally, one can use the on-site Kerr and cross-Kerr nonlinearities to simulate spin lattices in the Boltzmann (weak-coupling) regime. Here the cross-Kerr gives the $\sigma_z \otimes \sigma_z$ spin-spin interaction, and the linear coupling gives the appropriate weak-coupling (RWA) version of $\sigma_x \otimes \sigma_x$. This would allow one to create both integrable and non-integrable spin lattices, and investigate the behavior of thermalization and typicality as a function of lattice size~\cite{Dubey12}, as well as the integrability/non-integrability transition~\cite{Olshanii12}.   

The configuration we have described above can also be used for quantum computing where the modes of a single resonator are the qubits or qdits. An advantage of this scenario is the ability to couple any qubit to any other qubit via the driving fields. The Kerr non-linearity, if sufficiently strong, allows the lowest two Fock-states to be used as a single qubit, by separating them from the higher states. In this case all single-qubit gates can be obtained by linear driving and the Kerr nonlinearity, and the cross-Kerr provides a controlled phase gate, completing the universal set~\cite{Gottesman98}. Alternatively, the Kerr non-linearity, combined with the linear mode coupling and the ability to squeeze (achieved by parametric driving~\cite{Rugar91}) is universal for continuous variable computing~\cite{Lloyd99}. 


%
%
%
%
 
\textit{Acknowledgements:} KJ is partially supported by the NSF projects PHY-1005571, PHY-0902906, and PHY-1212413, and by the ARO MURI grant W911NF-11-1-0268. 
 


%

\end{document}